# Shear-induced particle diffusivities from numerical simulations

By MARCO MARCHIORO† AND ANDREAS ACRIVOS

The Levich Institute, T-1M, The City College of the City University of New York,

New York, NY 10031, USA

Using Stokesian dynamics simulations, we examine the flow of a monodisperse, neutrally buoyant, homogeneous suspension of non-Brownian solid spheres in simple shear, starting from a large number of independent hard-sphere distributions and ensemble averaging the results. We construct a novel method for computing the gradient diffusivity via simulations on a *homogeneous* suspension and, although our results are only approximate due to the small number of particles used in the simulations, we present here the first values of this important parameter, both along and normal to the plane of shear, which have ever been obtained directly either experimentally or numerically. We show furthermore that, although the system of equations describing the particle motions is deterministic, the particle displacements in the two directions normal to the bulk flow have Gaussian distributions with zero mean and, a variance which eventually grows linearly in time thereby establishing that the system of particles is diffusive. In addition we show that although the particle evolution equations are, in principle, reversible, the suspension has in fact a finite correlation time $T_c$ of the order of the inverse shear rate. For particle concentrations up to 45%, we compute the corresponding tracer diffusivities both from the slope of the mean square particle displacement as well as by integrating the cor-responding velocity autocorrelations and find good agreement between the two sets of results.

† Present address: RiskMap SpA, via Gian Battista Vico 4, Milan, Italy.

---

## 1. Introduction

It is generally accepted by now that shear-induced particle diffusion in a shear flow, under conditions of vanishingly small Reynolds numbers, plays a key role in the dynamic behavior of concentrated suspensions of non-Brownian particles (see Acrivos 1995). This phenomenon has been studied extensively, both experimentally as well as computationally (Breedveld *et al.* 1998; Bossis & Brady 1987), and several models have been proposed and then tested experimentally for determining the particle concentration profiles and flow characteristic of such suspensions under a variety of conditions (Leighton & Acrivos 1987*b*; Phillips *et al.* 1992; Nott & Brady 1994). Here we consider a monodisperse suspension of neutrally buoyant spheres in a simple shear flow and, by means of computer simulations, focus on the three basics aspects of this rapidly evolving field: 1) the statistics of the particle displacements; 2) the evaluation of the particle tracer diffusivity via both the evolution with time of the mean square particle displacement as well as from the integral of the velocity autocorrelation function; 3) the determination of the gradient diffusivity from the particle trajectories in a *homogeneous* suspension.

The aim of the first part of the paper is to try and answer once and for all the question which has been raised in the past by several investigators in this field, namely whether diffusion can possibly arise from a deterministic time evolution of the configuration of the system of particles in a space-filling suspension undergoing shear. To resolve this issue, we shall present the results of numerical simulations which show that, under a variety of conditions, a tracer particle diffuses in a homogeneous suspension being sheared even under the action of exclusively deterministic forces. In this contest, a system is said to be diffusive if any of its coordinates has a displacement, in time, that has a Gaussian distribution with zero mean, and a variance which, at least for large times, grows linearly with time. We shall also discuss briefly how macroscopic irreversibility can occur in a system which is reversible on a microscale.

The second part of the paper concerns the particle tracer diffusivity. Traditionally, see Bossis & Brady (1987), this coefficient for has been obtained starting from a single hard-sphere particle configuration and then computing particle displacements over time intervals which, even for a monolayer, were typically at least two orders of magnitude larger than $1/\gamma$, the characteristic time of the applied shear rate $\gamma$. Here, guided by the recent experiments of Breedveld *et al.* (1998) who evaluated the tracer diffusivity using a large number of independent displacement data taken over time intervals $1/\gamma$ or even lower, we performed simulations starting from a large number of independent hard-sphere particle configurations and then shearing the suspension for times $10/\gamma$ to $15/\gamma$, depending on the particle concentration, in order to achieve the equilibrium particle configurations pertaining to a homogeneous suspension in a simple shear. We then computed the diffusivity along and normal to the plane of shear from the ensemble average of the square displacement of the particles, as well as, apparently for the first time from numerical simulations, from the integral of the corresponding velocity autocorrelation function. The advantages of the ensemble average technique will be discussed, and it will be shown that there is close agreement between the values of the tracer diffusivity using the two methods just referred.

In the third part of this paper we present the first ever computed values of the gradient diffusivities, again along and normal to the plane of shear, which were obtained by applying a technique recently developed by Marchioro, Tanksley & Prosperetti (2000*a*,*b*), for the purpose of constraining closure relations in suspensions. The great advantage of this technique is that all the computations are performed on a *homogeneous* suspension. The values for the gradient diffusivity thus computed are in good agreement with the few available experimental data considering the fact that the maximum number of spheres used in our simulations was only 64.

## 2. Simulation method

The calculation of the diffusion coefficients is based on the numerical simulation of a suspension of spheres in a simple shear flow under Stokes flow conditions, and in the absence of Brownian and buoyancy effects. Although, in principle, any simulations method would do, we have chosen to use the well-known method of Stokesian dynamics of Brady & Bossis (1988) which has been developed specifically for computing the hydrodynamic interactions among an infinite suspension of particles at zero Reynolds numbers.

As was shown already by Brady & Bossis (1985), adjacent spheres interacting purely through hydrodynamic forces spend a great deal of time almost in contact with one another with the corresponding separation distances being less than $10^{-8}$ sphere radii. Also Dratler & Schowalter (1996) (see also Melrose & Ball 1995) found that, if the non-overlapping condition was strictly enforced, this minimum separation distance depended on the time step of integration, in the sense that adjacent spheres would come closer and closer together as this time step was decreased. All the authors referred above also concluded, however, that such simulations could still be performed provided a small repulsive force was introduced between adjacent pairs of particles in very close proximity to each other. The simulation results were then found to be independent of the detailed expression of this force provided it was sufficiently short ranged. The introduction of such a force actually models, at least qualitatively, the behavior of real physical systems in that, in the gaps between the particles, the presence of residual Brownian forces and/or the roughness of the spheres always play a central role.

For this inter-particle force we used the following repulsive force between two spheres, already well-tested in Stokesian dynamics simulations (Brady & Bossis 1985),

$$\mathbf{F}_{\alpha\beta} = F_0 \frac{\lambda e^{-\lambda\varepsilon}}{1 - e^{-\lambda\varepsilon}} \mathbf{e}_{\alpha\beta}, \qquad (2.1)$$

where, $6\pi\mu a^2 \gamma \mathbf{F}_{\alpha\beta}$, with $\mu$ being the viscosity of the suspending liquid and $a$ the sphere radius, is the force exerted on sphere $\alpha$ by sphere $\beta$, $F_0$ is a dimensionless coefficient reflecting the magnitude of this force, $\lambda$ is related to the range of the force, $\varepsilon$ is the distance of closest approach between the surfaces of the two spheres divided by $a$, and $\mathbf{e}_{\alpha\beta}$ is the unit vector connecting their centers pointing from $\beta$ to $\alpha$. In the simulations reported here, $\lambda F_0$ was chosen to be unity and $\lambda = 1000$. Simulations using other values for these parameters, namely $\lambda = 10^4, 10^5, 10^6$ and $F_0 = 1, 10^{-1}, 10^{-2}$, gave results which differed from one another by, at most, 5%.

The detailed description of the simulation method can be found elsewhere, see e.g. Brady & Bossis (1988), hence we provide here only a brief overview. We consider the zero Reynolds number flow of a suspension of hard spheres in simple shear characterized by a shear rate $\gamma$. The spheres are placed in a periodic box, typically a cube with dimensions uniquely determined by the number of spheres $N$ and the volume fraction $\phi$, and periodic boundary conditions are applied. The volume of the box $V$, the number of spheres $N$ and the volume fractions $\phi$ are related by the expression

$$\phi = \frac{4\pi N a^3}{3V}, \qquad (2.2)$$

where $a$ is the radius of the spheres. Let $x$, $y$, and $z$ be, respectively, the directions of the imposed shear flow, the velocity gradient, and the vorticity, and let $v$ and $w$ be the components of the particle velocity in the $y$ and $z$ directions, respectively. If $\mathbf{x}_N$ denotes the $3N$ dimensional vector of the position of the particles, the evolution of the particle trajectory can be represented by

$$\delta \mathbf{x}_N = \delta \mathbf{x}^a + \delta \mathbf{x}^H + \delta \mathbf{x}^F, \qquad (2.3)$$

where $\delta \mathbf{x}^a$ is the affine displacement in the $x$ direction due to the imposed shear, $\delta \mathbf{x}^H$ is the displacement due to pure hydrodynamic interactions, and $\delta \mathbf{x}^F$ is the displacement due to the inter-particle force defined in (2.1). A large number $N_c$, typically a few hundred, of simulations were performed starting from different initial hard-sphere configurations.

The computational cost of calculating both the many-body far-field interactions and the pairwise-additive lubrication forces between the particles is $O(N^3)$ for a single configuration, where $N$ is the number of spheres in the unit cell. In our case, this limited the number of particles in the unit cell to be not greater than 64.

In what follows, all the variables have been rendered dimensionless using the radius of the spheres $a$ as the characteristic length scale and $\gamma^{-1}$, the inverse of the imposed strain rate, as the characteristic time. As a consequence the diffusivities are scaled by $a^2\gamma$.

## 3. Statistics of the particle displacement

There exists experimental evidence that, even at vanishingly small Reynolds numbers and seemingly negligible Brownian force effects, particles in a concentrated suspension exhibit diffusive behavior (Leighton & Acrivos 1987$a$). More recently, it was also shown by Breedveld *et al.* (1998) that, in a sheared suspension at effectively infinite Peclet numbers, the displacements of tagged particles have an approximately Gaussian distribution even when evaluated for small time intervals.

In a physical experiment, however, particles are not perfectly spherical and a residual Brownian motion as well as minute inertia effects are always present within the suspension, hence it appears reasonable to suppose that non-deterministic irreversible forces cannot be completely eliminated under even the most idealized experimental conditions. The advantage of numerical simulations, on the other hand, is that it is possible to control more precisely all the parameters that affect the flow and, at least in principle, eliminate completely all non-deterministic and irreversible forces.

The first goal of the present paper is to show, from Stokesian dynamics simulations, that particle diffusion in a sheared suspension can take place even under the action of exclusively deterministic forces. This should not come as a surprise giving that, as is well known, even low-dimensional deterministic dynamical systems can exhibit diffusion (see Schuster 1989, page 32) which is usually taken as a sign of chaotic behaviour. Although we have not been able to prove that the system of equations describing the evolution of the particle positions is chaotic, we suggest that this is a reasonable hypothesis and indeed the cause of the phenomenon of shear induced diffusion. To this end we shall first examine the statistical properties of the displacement of a test particle in a sheared suspension.

Consider then $N$ spherical particles in the unit box flowing under shear. Denoting by $y(\tau)$ the $y$ coordinate of a generic particle at time $\tau$, we focus on the statistics of

$$\Delta y = y(\tau + t) - y(\tau), \qquad (3.1)$$

where $\tau$ is taken to be large enough to ensure that there is no dependence from the initial hard-spheres particle distribution (in our simulations $\tau \simeq 5$ is sufficient). The determination of the statistical properties of $\Delta y$, i.e. its *pdf*, requires a large number of samples.

Since, at present, Stokesian dynamics doesn't allow us to simulate systems with more than $N = 64$ particles, we rely on data for $\Delta y$ as obtained from a large number $N_c$ of different simulations starting with $N_c$ independent hard-sphere configurations. In this way the number of samples available for the determination of the statistics of $\Delta y$ is enlarged to $N \times N_c$ (since the computational cost is $O(N^3 N_c)$, it is convenient to choose $N$ small and $N_c$ large). Alternatively, a large number of different (although not independent) $\Delta y$'s could be taken from different time slices of a single configuration running for a long time. We checked that both approaches give the same *pdf*.

Figure 1 shows the *pdf* of (a) $\Delta y$ and (b) $\Delta z$ as obtained with $N = 16$ and $N_c = 512$ at a volume fraction $\phi = 25\%$, for different $t$ and for $\tau=5$, together with the plot of the Gaussian distributions with the same variance as the numerical data. It is evident to the eye that the distributions of both $\Delta y$ and $\Delta z$ are approximately Gaussian and, indeed, the calculations of higher moments confirm that the Gaussian distribution is approached for increasing $t$. A similar distribution for the displacements of the particle positions was observed experimentally by Breedveld *et al.* (1998).

The fact that the displacements have a Gaussian distribution suggests that some randomness entered the system, but does not mean as yet that the particles actually diffuse. To prove diffusion, it is necessary to show that the variance of the particle displacements grows linearly in time. To this end let us define

$$\xi(t,\tau) = \frac{1}{2N} \sum_{\alpha=1}^{N} [y^\alpha(\tau + t) - y^\alpha(\tau)]^2 , \qquad (3.2)$$

where $y^\alpha(t)$ is the $y$-coordinate of particle $\alpha$ at time $t$ starting from some initial hard-sphere distribution. If diffusion is indeed present in such a system, then $\xi$ should eventually grow linearly with $t$ with the diffusion coefficient being equal to the slope of this curve. Figure 2 shows a plot of $\xi(t,0)$ for $N = 27$, $\phi=35\%$, up to $t = 1600$. It seems reasonable, but certainly questionable, to conclude that $\xi(t,0)$ grows linearly in time. A more compelling proof of the existence of diffusion will be given later on in section 4.3.

Assuming, for the time being, that $\xi(t,0)$ is linear in $t$, we face the problem of determining its exact slope. This difficulty is illustrated in Figure 2 which shows linear least-square fits evaluated in two different regions of the time domain. The resulting slopes are a quite different, and the exact value of the diffusion coefficient seems to depend on the interval chosen to interpolate a linear fit.

This ambiguity is due to the statistical nature of $\Delta y$ and, in statistical methods, is generally encountered in evaluating the variance of certain random variables. Specifically, suppose that $\psi_1 \ldots \psi_n$ are $n$ independent Gaussian random variables with zero mean and unknown variance $\sigma^2$. The latter can be estimated by the random variable $S_n$, defined as

$$S_n = \frac{1}{n} \sum_{i=1}^{n} \psi_i^2 \,. \tag{3.3}$$

Since $\langle S_n \rangle = \sigma^2$ and, from the law of large numbers, $\lim_{n \to \infty} S_n = \sigma^2$, $S_n$ is a good estimate for the variance of $\psi_i$. However, $S_n$ is not equal to $\sigma^2$ for finite $n$ and the error of this estimate, i.e. the *r.m.s* of $S_n$, can be shown to be (Folland 1984, page 293)

$$\mathrm{Var}(S_n) = \langle (S_n - \langle S_n \rangle)^2 \rangle = 2 \frac{\sigma^4}{n} \,. \tag{3.4}$$

In the case under consideration, we identify the $\psi_i$'s with $\Delta y$ even though, strictly speaking, the particle displacements for a single configuration are generally not independent, and note that the variance $\sigma^2$ of the *pdf* of $\Delta y$ grows with $t$. In order to minimize the error in evaluating $\sigma^2$ one should therefore strive to have both a small $t$ and a large number of degrees of freedom $n$. Since a single configuration gives, at most, $n=N$, we can increase $n$ using a large number $N_c$ of configurations so that $n=N \times N_c$. Clearly, then, in view of the arguments given above, one should perform a large number of short simulations starting from a large number of independent configurations. Again, an alternate approach would be to take the different configurations by dividing one single long run into non overlapping intervals. This method was tried in few cases and gave results equivalent to those using the method of ensemble averaging.

Although all our computations were performed in the presence of the interparticle force given by (2.1), there exists the fundamental issue of whether diffusion should still occur if this force were absent. Recall that, under the action of exclusively hydrodynamic effects, the particle configuration in the sheared suspension would evolve according to the system of the Stokes equations which are reversible as well as deterministic. The latter implies, of course, that, given the initial particle configuration at time $\tau$, it should be possible, at least in principle, to determine *exactly* the positions of all the particles at time $\tau + T$. On the other hand, reversibility implies that, upon reversing the direction of shear at time $\tau + T$ one should be able to return these same particles to their initial positions at time $\tau + 2T$ giving $\xi(2T, \tau) = 0$, where $\xi$ is defined by (3.2). Since such a result is incompatible with diffusion, one is left with the task of explaining how a macroscopic irreversibility, such as diffusion, could possibly arise in a system which is reversible on a microscale.

An analogous paradox has been known for a long time in classical statistical mechanics where the movement of a large number of molecules whose equations of motions are reversible give rise to irreversible phenomena that are described by irreversible macroscopic equations such as the Boltzmann equation (see Kreuzer 1981, for an extensive review).

In our case, the result $\xi(2T, \tau) = 0$ referred to above presupposes that the positions of the particles at time $\tau + T$, i.e. at the instant where the direction of the shear is reversed, is known with infinite precision. But, if for some reason, this piece of information is slightly imprecise, then the initial positions of the particles cannot be recovered. In an experiment this loss of information arises from unavoidable small irreversible effects, i.e. surface roughness, Brownian and inertial forces etc., which can never be eliminated, while, in computer simulations, it enters through round off errors and imperfect computations.

As an illustration, consider the evolution of $\langle \xi(t, \tau) \rangle$, shown in Figure 3, with $F_0=10$, $\tau = 4$, and $\lambda = 10^3, 10^4, 10^5$, where the brackets denote the ensemble average over $N_c=512$ initial hard-sphere configurations ( We have chosen to illustrate the evolution of $\langle \xi(t,\tau) \rangle$ rather than that of $\xi(t,\tau)$ since the latter is both irregular and highly dependent on the particular initial hard-sphere configuration; hence the issue of reversibility cannot be probed as clearly with $\xi(t,\tau)$ as it can by examining the response of $\langle \xi(t,\tau) \rangle$ to flow reversal.) It is seen that if the direction of shear is reversed at time $\tau + T = 10$, the computed values of $\langle \xi(t,\tau) \rangle$ tend to be be symmetric about $\tau + T = 10$ for a short time beyond this point but eventually resume growing with increasing $t$. The fact that the minimum of $\langle \xi(2T,\tau) \rangle$, which was reached after a time $\tau + T = 13$, was only slightly affected by choosing three very different values for the strength of the interparticle force suggests that any small departure from reversibility will amplify at least exponentially. We conclude therefore that, upon reversal of the direction of shear, the suspension *loses its memory* after a time interval of the order of one strain, the value of which depends primarily on the particle concentration and the details of the inter-particle force, and that thereafter the evolution of the particle configuration proceeds as if the flow reversal had never taken place.

## 4. Autocorrelation time and the self-diffusion coefficients

In the previous section we showed that the *pdf* of the particle displacements in the two transverse directions is Gaussian but that, irrespective of the length of the simulation, it is not possible to determine, from equation (3.2) starting from a single hard-sphere distribution, an unambiguous linear relationship between their average square displacements and the time $t$. We also deduced from equation (3.4) that, with simulations using a small number of particles, it is necessary to employ data from different configurations in order to determine the diffusion coefficients with reasonable accuracy. This approach was suggested first by Breedveld *et al.* (1998) who determined experimentally the diffusion coefficients from the *pdf* of the particle displacements as obtained from many different configurations, and was followed by Foss & Brady (1999) in their computation via numerical simulations of the diffusion coefficient in the presence of Brownian forces. More precisely, let us define

$$D_{yy}(t,\tau) = \frac{1}{2}\frac{\mathrm{d}}{\mathrm{d}t}\langle[y(\tau+t)-y(\tau)]^2\rangle, \quad (4.1)$$

where

$$\langle[y(\tau+t)-y(\tau)]^2\rangle = \frac{1}{N_c}\sum_{i=1}^{N_c}\frac{1}{N}\sum_{\alpha=1}^{N}[y_i^\alpha(\tau+t)-y_i^\alpha(\tau)]^2, \quad (4.2)$$

with $\tau$ being non negative, and $y_i^\alpha(t)$ the $y$-coordinate of particle $\alpha$ at time $t$ taken from configuration $i$. An analogous definition for $D_{zz}(t,\tau)$ can be obtained by substituting $y$ with $z$ in equations (4.1) and (4.2).

In order to have diffusion in the sheared suspension, it is necessary that the mean square displacements in the $y$ and in the $z$ directions both grow linearly in time, in which case the long time particle self-diffusion coefficients $\hat{D}_{yy}$ in the $y$-direction and $\hat{D}_{zz}$ in the $z$-direction, are defined as the limit, respectively, of $D_{yy}(t,\tau)$ and $D_{zz}(t,\tau)$ when $t\to\infty$, these limiting values being independent of $\tau$.

Figure 4(a) shows a plot of $D_{yy}(t,0)$ for $N_c = 512$, $N = 27$, and $\phi = 25\%$, $35\%$, and $45\%$, from where it is clear that $D_{yy}(t,0)$ does not reach a constant value for increasing $t$ but oscillates with a period equal to precisely one for a cubic box. The data shown in figure 4(a) were obtained using an ensemble of initial particle configurations as generated by a hard-sphere molecular-dynamic method, in which all the particles are placed in the same cubic box but in different locations avoiding overlapping.

The presence of these oscillations prevent us from evaluating the limit of $D_{yy}(t,0)$, and hence the diffusivity, except in an approximate way. In order to find the cause of these oscillations, and thereby to devise a method for eliminating them, we next consider the simpler case of evaluating the intensity of the velocity fluctuations.

### 4.1. *Oscillations of the intensity of the velocity fluctuations*

Consider the intensity of the velocity fluctuations in the $y$ direction defined as,

$$\langle v^2(\tau) \rangle = \frac{1}{N_c} \sum_{i=1}^{N_c} \frac{1}{N} \sum_{\alpha=1}^{N} [v_i^\alpha(\tau)]^2 \ , \qquad \text{with } \langle v(\tau) \rangle = 0, \tag{4.3}$$

where $v_i^\alpha(\tau)$ is the $y$-component of the velocity of particle $\alpha$ at time $\tau$ starting from hard-sphere configuration $i$. A plot of $\langle v^2(\tau) \rangle$ for $N_c = 512$, $N = 27$, and $\phi = 25\%, 35\%,$ and $45\%$, is shown in figure 5(a), from where it is clearly visible that $\langle v^2(\tau) \rangle$ oscillates with exactly the same period as did $D_{yy}(t,0)$. A shown in figure 5(b), however such regular oscillations were not observed in the evolution of the intensity of the $z$ component of the velocity fluctuations $\langle w^2(\tau) \rangle$.

To study more quantitatively this effect in the long-time behavior, we considered the case of the evolution of a single configuration sheared for a very long time ($\tau_\infty = 6554$). To this end we evaluated the Fourier transform of $\langle v^2(\tau) \rangle$,

$$F_{\langle vv \rangle}(\nu) = \left| \int_0^{\tau_\infty} \mathrm{d}\tau \ \langle v^2(\tau) \rangle \ e^{2\pi i \nu \tau} \right|, \tag{4.4}$$

shown in figure 6 as a function of the frequency $\nu$. Two very sharp peaks are clearly noticeable at $\nu = 0$ and at $\nu = 1$, and a smaller one at $\nu = 2$. The peak at $\nu = 0$ is due to the constant part of the signal and is not related to the oscillations, but the remaining peaks, in particular that at $\nu = 1$, refer to the anomaly under discussion. For an initially cubic shape, the simulations were repeated for different number of particles and different sizes of the box but the locations of the peaks remained unchanged. This procedure was then repeated for four different shapes of the initial box. The results are reported in table 1 for a monolayer of particles with areal fraction equal of 50%, where the horizontal length and height of the boxes, denoted by $L$ and $H$ respectively, is listed in the first column, the location of the highest peak (not considering that at $\nu = 0$) in the second, and the magnitude of that peak in the third. Clearly there is a direct relation between the side-height ratio of the initial box and the location of the maximum peak.

It was also noticed that the regular oscillations slowly decrease in magnitude as the height of the cubic cell, hereby denoted by $H$, is increased. In order to check this dependence more quantitatively, a simulation was performed with $\phi$=45% and $N$=64 giving a side length 4/3 times bigger than that of the cubic box with 27 particles at the same volume fraction. It was found that the intensity of the oscillations relative to the mean value of $v^2$ was reduced from 0.32 to about 0.24, the ratio being 4/3. By assuming an inverse relation between the magnitude of the oscillations and $H$, we therefore conclude that to reduce the oscillations to values less than 5% of the mean it would be necessary to increase the cell side to about 6.5 times that of the box containing 27 particles. At $\phi$=45%, this would then require a box with more than 7000 particles, clearly far in excess of the number of particles which can be handled using presently available computers capabilities.

The dependence of the amplitude of the regular oscillations on $H$ might suggest that this phenomenon could be due to the long-range hydrodynamic interactions of each particle with its images in the lattice. This hypothesis was disproved, however, when we performed simulations of the sheared suspension by neglecting the long range interactions. This was accomplished in two independent ways: by running simulations with a code based on the method describe by Ball & Melrose (1997), as well as by employing a variation of the Stokesian dynamics code in which only lubrication forces were retained. In both cases the frequency of the oscillations remained unchanged and their magnitude was comparable with that obtained from simulations in which the complete hydrodynamic interactions were retained. From these results we deduced that the regular oscillations are not due to the velocity field induced by a sphere on its images, but rather by the presence of an uninterrupted chain of particles, interacting through lubrication forces, which span a major fraction of the length of the cell.

The regular oscillations in the intensity of the velocity fluctuations seem, therefore, to be due to a finite-size effect which, apparently, could be eliminated, for any given configuration, only by increasing the size of the box. As we shall presently show, however, the same goal can be accomplished by starting the computations with an ensemble of configurations that are randomly out of phase with each other.

### 4.2. *Elimination of the regular oscillations*

To understand the nature of the regular oscillations one has to consider the details of the numerical simulations from which they arise. To ensure the correct representation of long range interaction with a finite number of particles, Brady *et al.* (1988) have shown that it is necessary to confine the particles in a periodic lattice. At this point, the only way to apply a shear in a dynamic way is by means of the Lees-Edwards boundary conditions, see e.g. Allen & Tildesley (1987), in which the shape of the lattice changes and deforms in time until it reaches the original cubic lattice shape(that is with exactly the same frequency as that indicated in table 1). Consequently, a spurious periodicity is introduced into the hydrodynamic calculations by the instantaneous shape of the lattice. On the other hand, the shape of the box in the $z$ direction is not altered, a fact which probably accounts for the lack of any observed regular oscillations in $\langle w^2(\tau) \rangle$.

Let us decompose the instantaneous velocity of a single particle in a certain configuration into a *physical* part, which we wish to measure, and an oscillating part,

$$v(\tau) = A(\tau) + B\, e^{2i\pi\nu\tau}, \tag{4.5}$$

where, $\langle A(\tau) \rangle = 0$, $\langle B \rangle = 0$, and $\nu$ is the frequency of the spurious oscillation. Since higher harmonics are, in general, very small we consider only the fundamental harmonic, in our case $\nu = 1$ for the cubic box. The expression for the intensity of the velocity fluctuations becomes then,

$$\langle v^2(\tau) \rangle = \langle A^2(\tau) \rangle + 2 \langle A(\tau)B \rangle \, e^{2i\pi\nu\tau} + \langle B^2 \rangle \, e^{4i\pi\nu\tau} \, . \tag{4.6}$$

From the graph in figure 6, we can estimate that, for that particular case, $\langle A^2 \rangle \simeq 1$, $\langle AB \rangle \simeq 0.01$, and $\langle B^2 \rangle \simeq 0.002$, hence we deduce that there exists a certain degree of correlation between $A$ and $B$. In order to eliminate the regular oscillations, we therefore introduce a random phase in the time of each single configuration. This can be accomplished by starting each simulation from a random negative time, so that equation (4.5) now becomes

$$v(\tau) = A(\tau) + B \, e^{2i\pi\nu(\tau+\eta)} \, , \tag{4.7}$$

where $\eta$ is random number ranging from 0 to 1 independent from the initial configurations. The expression for the velocity fluctuations then reads

$$\langle v^2(\tau) \rangle = \langle A^2(\tau) \rangle + 2 \langle A(\tau)B \rangle \, e^{2i\pi\nu\tau} \langle e^{2i\pi\nu\eta} \rangle + \langle B^2 \rangle \, e^{4i\pi\nu\tau} \langle e^{4i\pi\nu\eta} \rangle = \langle A^2(\tau) \rangle \, , \tag{4.8}$$

which no longer contains the effects of the regular oscillations given that $\langle e^{2i\,\ell\,\pi\nu\eta} \rangle = 0$, when $\ell$ is an integer. This procedure is equivalent to changing the definition of the average from that of (4.3) to

$$\langle\langle v^2(\tau) \rangle\rangle \; = \frac{1}{N_c} \sum_{i=1}^{N_c} \frac{1}{N} \sum_{\alpha=1}^{N} [v_i^\alpha(\tau + \eta_i)]^2 \; , \tag{4.9}$$

where $\eta_i$ is a random number for each configuration uniformly distributed between 0 and 1. As shown in figure 7, which depicts a plot of $\langle\langle v^2(\tau) \rangle\rangle$ obtained using the procedure of randomized phases just described (see caption for details), the large regular oscillations have disappeared. Also, the remaining noise can be made as small as desired by increasing the number of configurations. We wish to remark parenthetically that the method which we have devised for the purpose of eliminating these regular oscillations is by no means unique in that any number of filtering techniques could have been employed for achieving the same goal. Our method has the great advantage, however, that it can be applied simultaneously to the computation of all the physical quantities of interest, such as the displacements, the velocities, and the velocity autocorrelation functions. We also wish to point out that if the regular oscillations seen in figures 4 and 5 were physical, e.g. due to the formation and destruction of clusters, rather than spurious as we have assumed, it would not have been possible to remove them just by shifting the origin of time in the initial hard-spheres configurations.

In what follows, we shall use the definitions of the averages similar to (4.9) and proceed to describe the calculation of the different coefficients characterizing the sheared suspension.

### 4.3. *Numerical results for the self-diffusivities*

We now return to the problem of establishing the diffusive behavior of the sheared suspension. To avoid the oscillations of figure 4, we evaluated $D_{yy}(t, \tau)$ from equation (4.1) using definition

$$\langle\langle [y(\tau + t) - y(\tau)]^2 \rangle\rangle = \frac{1}{N_c} \sum_{i=1}^{N_c} \frac{1}{N} \sum_{\alpha=1}^{N} [y_i^\alpha(\tau + t + \eta_i) - y_i^\alpha(\tau + \eta_i)]^2 , \qquad (4.10)$$

for the average displacement, instead of definition (4.2). An analogous definition holds for the average square displacements $D_{zz}(t, \tau)$ in the $z$ direction. A plot of $D_{yy}(t, 0)$ as obtained using both definitions (4.10) and (4.2), for 27 particles, with $N_c$=512, $\phi$=15%, is given in figure 8. In the limit of large $t$, in the particular case of this figure for $t \geq 8$, $D_{yy}(t, 0)$, as obtained from (4.1) and (4.10), approaches a constant value as $t \to \infty$ so that, in contrast to using (4.2), the self-diffusion coefficient $\hat{D}_{yy}$ can now be determined.

Analogous result applies for $\hat{D}_{zz}$ which, however, can be computed using either (4.2) or (4.10), since there are no regular oscillations in the $z$ direction.

In table 2, we report the computed values of the self-diffusivity both in the $y$ and $z$ directions, for a volume fraction $\phi = 35\%$, as obtained with a different number of particles in the unit cell, $N = 16, 27, 54$, and 64 with the reported error being the $r.m.s.$ obtained from the ensemble average. Clearly, the computed self-diffusion coefficients are strongly dependent on the number of particles used in the simulations. We therefore plotted $\hat{D}_{yy}$ and $\hat{D}_{zz}$ as functions of the inverse box height $H$, with $1/H$ ranging between 0.1 and 0.2, and observed that, with a correlation coefficient always greater than 0.96, the points fall on a straight line so that a linear extrapolation to $1/H = 0$ appeared to be permissible. Figure 9 shows a comparison of the values of (a) $\hat{D}_{yy}$ and (b) $\hat{D}_{zz}$, extrapolated to $1/H = 0$, with previous experimental and numerical results. For volume fractions below 30% there is a fairly good agreement with the data in the literature. For more concentrated suspensions, however, such a comparison is more difficult to make because of the scatter in the experimental points. Even so, the results of this work are closer to the experimental values than those of Foss & Brady (1999) whose simulations were performed with only 27 particles and at a Peclet number $Pe = 10^3$. Also, consistent with the experimental data, our computed values of $\hat{D}_{yy}$ are significantly higher than those of $\hat{D}_{zz}$ at the same volume fraction. Several discrepancies are clearly evident in figure 9, however, which require further study.

### 4.4. Numerical results for the autocorrelation function

We have already observed that the reversibility of the flow manifests itself only for times of the order of one strain and that the diffusion equation can be used to describe the macroscopic system for times that are much larger than that. In this subsection we proceed to investigate in more detail this transition between the reversible flow at short time scales and the irreversible flow at large times.

To this end we consider the correlation between the transverse velocity of a certain particle at two different times, also known as the velocity autocorrelation function. The evaluation of this function for the velocity in the $y$ direction, using the random-phase adjustment to avoid the regular oscillations, is given by

$$\langle\langle v(\tau+t)v(\tau)\rangle\rangle = \frac{1}{N_c}\sum_{i=1}^{N_c}\frac{1}{N}\sum_{\alpha=1}^{N} v_i^\alpha(t+\tau+\eta_i)\,v_i^\alpha(\tau+\eta_i)\ , \qquad (4.11)$$

where $v_i^\alpha(t)$ is the $y$-component of the velocity of particle $\alpha$ at time $t$ in the configuration $i$, and $\eta_i$ is a random number between 0 and 1; whilst the autocorrelation function in the $z$ direction has a similar definition with $v$ replaced by $w$. A typical graph of $\langle\langle v(\tau+t)v(\tau)\rangle\rangle$, normalized by $\langle\langle v^2(\tau)\rangle\rangle$, is shown in figure 10 (a), for $\tau=5$, $N_c=256$, $N=27$, $\phi=35$, and $\phi=45\%$. We notice that the autocorrelation function quickly drops to zero after about 0.8 strains, remains negative for about four strains, and then becomes negligible shortly thereafter. A similar qualitative behaviour is obtained at different volume fractions or with a different number of particles. The presence of a long tail in $\langle\langle v(\tau+t)v(\tau)\rangle\rangle$ suggests that, even though the velocity auto-correlation becomes very small (less than 20%) within one strain, a much longer time is required, about four or five strains for the particular case of figure 11, for the suspension to evolve into a completely independent state. The autocorrelation function for the velocity in the $z$ direction, shown in figure 10 (b), has a similar qualitative behaviour.

Knowledge of the autocorrelation function also allow us to calculate the self-diffusion coefficient in an independent way using the well-known expression (for details, see Berne & Pecora 1976, page 86)

$$\frac{1}{2}\frac{\mathrm{d}}{\mathrm{d}t}\langle\langle[y(\tau+t)-y(\tau)]^2\rangle\rangle = \int_0^t \langle\langle v(\tau+t')v(\tau)\rangle\rangle\,\mathrm{d}t'\ . \qquad (4.12)$$

Therefore,

$$\tilde{D}_{yy} = \int_0^\infty \lim_{\tau \to \infty} \langle\langle v(\tau+t)v(\tau)\rangle\rangle \, dt, \tag{4.13}$$

and similarly for the coefficient $\tilde{D}_{zz}$ in the $z$-direction. Values for $\tilde{D}_{yy}$ and $\tilde{D}_{zz}$ for $\phi = 35\%$, as obtained from equation (4.13) with $N = 16, 27, 54,$ and 64, are reported in table 2, where the indicated error is the *r.m.s.* of the ensemble average. The values of the diffusivities from the average square displacements and from the integral of the autocorrelation function are also essentially equal with each other for the other volume fractions considered but not printed. We notice that, for the same number of configurations, the error in the evaluation of $\hat{D}_{yy}$ and $\hat{D}_{zz}$, is consistently smaller than that of $\tilde{D}_{yy}$ and $\tilde{D}_{zz}$.

Shown in figure 11 is a comparison between $\int_0^t \langle\langle v(\tau+t')v(\tau)\rangle\rangle dt'$ (plotted as solid line) and $D_{yy}(t)$, dashed line from equations (4.1) and (4.10), which are seen to be equal within numerical error. This same degree of agreement was also found in all the other cases which were considered. An analogous comparison for the components in the $z$ direction is also shown in figure 12, where the curves are seen to be very similar to those in figure 11.

The form of $\langle\langle v(\tau+t)v(\tau)\rangle\rangle$ and the relation (4.12) just verified allow us to understand more fully the dependence of $\frac{1}{2}\langle\langle [y(\tau+t) - y(\tau)]^2\rangle\rangle$ on $t$ (plotted as a dotted line in figures 11 and 12). For $t \to 0$, this quantity grows quadratically,

$$\frac{1}{2}\langle\langle [y(\tau+t) - y(\tau)]^2\rangle\rangle \simeq \langle\langle v^2(\tau)\rangle\rangle \, t^2 \qquad \text{as} \qquad t \to 0, \tag{4.14}$$

as confirmed by the numerical results. Thus, at these time scales the motion is not diffusive and completely reversible (as also shown by figure 3). This constitutes a very important difference between the suspension considered here, in which Brownian motion is neglected, and a suspension at finite Peclet numbers where the short-time diffusivities are independent of $t$ and proportional to the mobility matrix (see, e.g., Brady 1994).

Moreover, in our case, an apparently linear regime is reached in the range $0.5 < t < 2$ due to the fact that the autocorrelation function changes sign at about $t=1$ where its integral reaches a local extremum. This led some earlier researchers, e.g. Foss & Brady (1999), to terminate their simulations at $t = 2$ in the mistaken belief that they had reached the asymptote for the diffusion coefficient. Finally, for $t > 5$ we have $\langle\langle v(\tau+t)v(\tau)\rangle\rangle \simeq 0$, so that its integral, $D_{yy}(t)$, becomes constant and the final diffusive steady state is reached. More precisely, let us define the *correlation time* as the smallest time $T_c$ such that

$$\frac{|\langle\langle v(\tau+t)v(\tau)\rangle\rangle|}{\langle\langle v^2(\tau)\rangle\rangle} \leq 0.05 \qquad \text{for any} \qquad t \geq T_c. \tag{4.15}$$

This correlation time is an important time scale for the flow of the suspension, in that it refers to the time interval which must be exceeded before the average particle displacement can be described with a diffusion equation. In table 3 we report the correlation times found from our numerical simulations for different number of particles in the unit cell and different values of the volume fraction $\phi$. Although the evaluation of the correlation time is evidently subject to large error, its order of magnitude at various volume fractions is of some importance. As expected intuitively, $T_c$ decreases with increasing volume fractions and its value does not seem to depend strongly on the number of particles used in the unit cell. Also, the fact that $T_c$ is significantly larger than unity especially at lower particle volume fractions, raises doubts regarding the accuracy of the diffusion coefficients reported by Breedveld *et al.* (1998) and by Breedveld (2000), and shown in figure 9, because, on account of the large scatter in their data, their measured mean square particle displacements are reliable only for values of the strain below about 0.7 and 2, respectively, both of which are below $T_c$.

In the previous section we showed how the displacements in the directions normal to the flow of the position of a tagged particle in a uniform sheared suspension have a Gaussian distribution, whereas, in this section, we showed that, for times greater than the correlation times reported in table 3, the average square displacements grow linearly with $t$ giving well defined diffusion coefficients. These results were obtained with simulations involving relatively small number of particles but, using ensemble averages. We wish to note that the main advantage of starting our simulations using initial hard-sphere configurations is that the sequence in which the suspension flow is computed for the different configurations is immaterial. This is very advantageous if a parallel machine or a cluster of machines is available. On the other hand, if only a single processor computer is available and the time required for $\langle\langle v^2(\tau)\rangle\rangle$ to reach its equilibrium steady state for a simple shear flow is larger then the correlation time, it may be preferable to obtain the different configurations of the ensemble by dividing a single simulation performed over a long time into many small parts and then proceeding as if these configurations were independent

## 5. Evaluation of the gradient diffusivity

Up to this point we have only dealt with suspensions that are homogeneous on a macroscale, and have determined some of the key parameters that result in the presence of simple shear, e.g. the shear-induced tracer diffusivities along and normal to the plane of shear, the corresponding velocity autocorrelations functions, the *correlation time* beyond which the suspension loses its memory etc. Most practical applications, however, involve suspensions in which the particle concentration profile is non-uniform with the result that a shear-induced particle flux is created from regions of high concentrations to low (Acrivos 1995). The key parameters here are the so called gradient diffusivities $D_\parallel$ and $D_\perp$ (Leighton & Acrivos 1987*b*), which appear in the linear relation between the particle flux **J** and the local particle concentration gradient in the two directions normal to the flow,

$$J_y = -D_\parallel \frac{\partial \phi}{\partial y} \qquad \text{and} \qquad J_z = -D_\perp \frac{\partial \phi}{\partial z}. \tag{5.1}$$

The evaluations of these two coefficients will be the subject of the remaining part of this paper.

### 5.1. *Non uniform suspension*

In principle, this gradient diffusivity in one of the transverse directions, say $y$, could be computed by artificially introducing into the suspension a finite force $F_y$, periodic in $y$ with period $2\pi/H$, and then determining via simulations how an initially statistically *homogeneous* suspension evolves under shear into a state in which the local particle concentration is also periodic in $y$. It can be shown that the relaxation time for this process equals $H^2/4\pi^2 D$ where $D$ is the corresponding gradient diffusivity. The difficulty with this approach is that the simulations must be performed on suspensions which are statistically non-homogeneous. Also, since the value of $D$ thereby obtained depends on the strength of this artificial force, the computations would have to be repeated for different values of this strength and the results extrapolated to zero.

In contrast, in the technique which we shall present below and which is a modification of that described by Marchioro *et al.* (2000 *a,b*), all the computations are performed on a suspension which is statistically homogeneous. The essence of this method is to create a small artificial inhomogeneity in the suspension, not by altering the particle trajectories in response to an applied force, but rather by changing the probability of the different configurations in a prescribed way according to the instantaneous state of inhomogeneity.

We begin by recalling that the local ensemble-average volume fraction of the particles at location $y$, is defined by

$$\phi(y,t) = \sum_{i=1}^{N_c} P(i,t) \sum_{\alpha=1}^{N} \frac{4\pi}{3} \delta(y - y_i^\alpha), \tag{5.2}$$

where $P(i,t)$, to be specified more precisely later, is the probability of the configuration labeled $i$ ($i=1, ..., N_c$), $4\pi/3$ is the volume of each sphere of unit radius, and $\delta$ is the Dirac delta function. An important point which must be kept in mind is that the specification of $P(i,t)$ for any given ensemble of configurations is arbitrary and that the specific expression chosen for $P(i,t)$ will determine the properties of the system and, according to the definition (5.2), the volume fraction profile $\phi(y,t)$ within the suspension.

Specifically, let us consider a particular choice of $P(i,t)$, namely $P_0(i,t)$, which gives rise to a homogeneous suspension where, in this contest, we say that the suspension is homogeneous with respect to $P$ in the $y$-direction if the value of $\phi(y,t)$ is the same for each $y$. Now if the $N_c$ initial configurations are sampled from a hard-sphere distribution, then there is no reason to give a higher or lower probability to any of these initial configurations, and hence, $P_0(i,0)=1/N_c$. We also established in section 3, however, that when the simple shear is applied to any of these initial configurations in the absence of an external force, the particles will be displaced in the $+$ or $-$ directions with equal probability (and similarly with respect to $z$). As a consequence, the different configurations will maintain the same relative probability and the choice $P_0(i,t)=1/N_c$ for every $t \geq 0$ will give rise to a uniform distribution.

To obtain a non uniform system it is therefore necessary to use a different definition of $P(i,t)$. Several choices are possible but a very convenient one (see, e.g. Marchioro *et al.* 2000*a*) is to define a probability $P_\varepsilon(i,t)$, according to

$$P_\varepsilon(i,t) = P_0(i,t)\left[1 + \varepsilon\,\sigma(i,t)\right], \tag{5.3}$$

where,

$$\sigma(i,t) = \sum_{\alpha=1}^{N} \sin k\,y_i^\alpha(t), \tag{5.4}$$

$\varepsilon$ is a small parameter which, as will be seen presently, cancels out of the analysis and $k = 2\pi/H$ is such that $\sin k\,y$ is periodic over the height of the box.

The volume fraction $\phi$, defined in (5.2) with $P$ replaced by $P_\varepsilon$, can be expanded in a Fourier series in the spatial coordinate to give

$$\phi(y,t) = \phi_0 \left[ 1 + S(t)\,\varepsilon \sin k\,y + C(t)\,\varepsilon \cos k\,y \right], \tag{5.5}$$

where $\phi_0 = 4\pi N/3V$, with $V$ being the volume of the cell, is the homogeneous volume fraction while

$$S(t) = \frac{2}{N} \sum_{i=1}^{N_c} P_0(i,t)\,[\sigma(i,t)]^2, \tag{5.6}$$

and

$$C(t) = \frac{2}{N} \sum_{i=1}^{N_c} P_0(i,t) \sum_{\alpha=1}^{N} \sin k\,y_i^\alpha(t) \sum_{\beta=1}^{N} \cos k\,y_i^\beta(t). \tag{5.7}$$

We shall presently show, however, that $C(t) = 0$ when the suspension described by $P_0$ is homogeneous and a sufficiently large number of configurations are used in the ensemble. To this end consider a possible configuration in the homogeneous ensemble, for example that labeled A in Figure 13. In the ensemble another configuration, B in our example, might be present which is obtained from A by mirror symmetry about the centerline. Because of symmetry,

$$\sin k\,y_A^\alpha = -\sin k\,y_B^\alpha, \qquad \cos k\,y_A^\alpha = \cos k\,y_B^\alpha, \tag{5.8}$$

for $\alpha = 1,\ldots,N$. But since the ensemble is homogeneous, configurations A and B will have the same probability, i.e. $P_0(i_A,t) = P_0(i_B,t)$, and therefore, from equation (5.7), $C(t)$ will vanish.

The ensemble average of the particle velocity in the $y$ direction is defined in a way similar to equation (5.2):

$$\overline{v(y,t)} = \frac{1}{\phi(y)} \sum_{i=1}^{N_c} P_\varepsilon(i,t) \sum_{\alpha=1}^{N} \frac{4\pi}{3}\,\delta(y - y_i^\alpha)\,v_i^\alpha(t), \tag{5.9}$$

where $v_i^\alpha$ is the velocity of particle $\alpha$ in the $y$ direction in configuration $i$. It is however more convenient to expand $\phi(y,t)\overline{v(y,t)}$, rather than $\overline{v(y,t)}$, in a Fourier series giving:

$$\phi(y,t)\overline{v(y,t)} = \phi_0 \left[ v^0(t) + v^s(t)\,\varepsilon\,\sin k\,y + v^c(t)\,\varepsilon\,\cos k\,y \right], \qquad (5.10)$$

where

$$v^0(t) = \sum_{i=1}^{N_c} P_0(i,t) \frac{1}{N} \sum_{\alpha=1}^{N} v_i^\alpha(t), \qquad (5.11)$$

$$v^s(t) = \sum_{i=1}^{N_c} P_0(i,t)\sigma(i,t) \frac{2}{N} \sum_{\alpha=1}^{N} v_i^\alpha(t) \sin k\,y_i^\alpha(t), \qquad (5.12)$$

$$v^c(t) = \sum_{i=1}^{N_c} P_0(i,t)\sigma(i,t) \frac{2}{N} \sum_{\alpha=1}^{N} v_i^\alpha(t) \cos k\,y_i^\alpha(t). \qquad (5.13)$$

The parameter $v^0(t)$ is simply the average velocity of the particles in the $y$-direction in the homogeneous suspension (i.e. when $\varepsilon = 0$) and therefore vanishes in a shear flow. To show that $v^s$ also vanishes, consider a possible configuration of the ensemble, for example configuration C of figure 14. Since the ensemble is homogeneous and representative of the whole configuration space, there exists another configuration, configuration D in the example of Figure 14, that can be obtained from C by a rotation of $\pi$ radians. As a consequence of the symmetry of the two configurations, for each $\alpha=1,\ldots,N$ we have

$$\sin k\,y_C^\alpha = -\sin k\,y_D^\alpha \qquad \text{and} \qquad v_C^\alpha = -v_D^\alpha. \qquad (5.14)$$

The two configurations also have the same probability in the homogeneous distribution and therefore $v^s = 0$ on account of equation (5.12). Note that the same argument does not lead to any conclusion regarding $v^c$ because

$$\cos k\,y_C^\alpha = \cos k\,y_D^\alpha. \qquad (5.15)$$

Next, we derive a relation between $S(t)$ and $v^c(t)$ given by equations (5.6) and (5.13), respectively, by taking the time derivative of $S(t)$, denoted by $\dot{S}(t)$, thereby obtaining

$$\dot{S}(t) = \sum_{i=1}^{N_c} P_0(i,t)\,2\,\sigma(i,t) \frac{2}{N} \sum_{\alpha=1}^{N} \frac{d}{dt} \sin k\,y_i^\alpha(t) = 2\,k\,v^c(t). \qquad (5.16)$$

4*Shear-induced particle diffusivities from numerical simulations*     27

Note that in the same way one could obtain $\dot{C} = 2\,k\,v^s$, showing that $v^s = 0$ from $C = 0$. Consequently, on making use of the above together with equations (5.5) and (5.10), we find that,

$$\phi(y,t) = \phi_0 + \phi_0\,S(t)\,\varepsilon \sin k\,y\,, \tag{5.17}$$

$$\overline{v(y,t)} = \frac{1}{2\,k}\dot{S}(t)\,\varepsilon\,\cos k\,y + O(\varepsilon^2)\,. \tag{5.18}$$

### 5.2. *Flux balance and numerical results*

Before proceeding with the non-stationary case, let us consider the system at steady state. After the suspension has been sheared for a long time, the concentration profile becomes

$$\phi_\infty(y) = \phi_0 + \phi_0\,S_\infty\,\varepsilon \sin k\,y\,, \tag{5.19}$$

where $S_\infty = \lim_{t\to\infty} S(t)$. To maintain this concentration profile which has a gradient in the $y$ direction, it is necessary to have a time-independent flux of particles $J_\infty$, that counterbalances the gradient diffusion flux thereby generated. Hence, in virtue of equation (5.1), this flux must satisfy

$$J_\infty = D_\parallel(\phi_\infty)\frac{\partial \phi_\infty}{\partial y} = \phi_0\,k\,D_\parallel(\phi_0)\,S_\infty\,\varepsilon\,\cos k\,y + O(\varepsilon^2)\,, \tag{5.20}$$

which also defines $D_\parallel$. In other words, the introduction of the artificial probability $P_\varepsilon(i,t)$ given by (5.3) and (5.4) is equivalent to introducing a flux along $y$ proportional to $\cos k\,y$ that, at steady state, counterbalances the gradient diffusion flux due to the non-uniform particle concentration which $P_\varepsilon(i,t)$ has created. Moreover, since the expression for $P_\varepsilon(i,t)$ does not depend explicitly on $t$, we shall retain (5.20) as representing this probability-induced flux for all $t \geq 0$.

In the general non-stationary case, the total flux of particles crossing each plane at

constant $y$ is given by

$$J^t \;=\; \phi\,\overline{v} \;=\; \frac{\phi_0}{2\,k}\dot{S}(t)\,\varepsilon\,\cos k\,y + O(\varepsilon^2)\,. \qquad (5.21)$$

This flux, is balanced by two terms; the shear-induced gradient flux given by

$$J^g \;\equiv\; -D_\|(\phi)\frac{\partial \phi}{\partial y} \;=\; -\phi_0\,k\,D_\|(\phi_0)\,S(t)\,\varepsilon\,\cos k\,y + O(\varepsilon^2)\,, \qquad (5.22)$$

and the flux $J_\infty$ defined in equation (5.20) created by $P_\varepsilon(i,t)$. Particle conservation requires that $J^t \;=\; J^g + J_\infty$, hence

$$\frac{1}{2\,k}\dot{S}(t) = -D_\|\,k\,[S(t) - S_\infty] + O(\varepsilon)\,. \qquad (5.23)$$

Notice that the term $\varepsilon\,\cos k\,y$ has disappeared from equation (5.23) the solution of which is

$$S(t) = S_\infty + [S(0) - S_\infty]\,e^{-2\,k^2\,D_\|\,t} \qquad \text{with } k = \frac{2\pi}{H}\,. \qquad (5.24)$$

By fitting the numerical results of $S(t)$ obtained from (5.6) to the form of equation (5.24), it is possible to extract the coefficient $D_\|$. Figure 15 shows the numerical values of $S(t)$ evaluated at a volume fraction $\phi=35\%$, with $N = 27$ particles, as well as the curve fit according to equation (5.24).

Before accepting this curve fit, however, it is necessary to check that these results do not depend on the particular choice of the initial distribution for the configurations used in the calculations. A good parameter characterizing the structure of the suspension at the initial time $t = 0$ is $S(0)$ which actually equals the static structure factor of the initial particle distribution at the lowest wave-number (for more details, see Berne & Pecora 1976, page 225). Different independent initial ensembles were therefore generated having different values of $S(0)$, and $S(t)$ was determined from simulations which were performed starting from these initial ensembles. The results which were obtained for the gradient diffusivity were found to be in agreement with each other within the measured error, i.e. the error found in fitting $S(t)$ using equation (5.24). As might have been anticipated, ensembles with higher values of $S(0)$ gave smaller errors in $D_\parallel$. For this reason ensembles with high $S(0)$ values, usually between 0.5 and 1.0, were used in subsequent simulations.

The gradient diffusivity in the direction normal to the plane of shear $D_\perp$, can be evaluated, using exactly the same method as for $D_\parallel$, by substituting the spatial coordinate $z$ for $y$ and the velocity component $w$ for $v$. The results for both $D_\parallel$ and $D_\perp$ are reported in table 4, for a volume fraction $\phi = 35\%$, as obtained with a different number of particles in each unit cell and a different number of configurations. Again, as in the case of the tracer diffusivities, these diffusion coefficients are strongly dependent on the number of particles in the unit cell and therefore, as before, a linear regression in $1/H$ was used to extrapolate the results to the limiting case of an infinitely large cell. In Figure 16 the gradient diffusion coefficients $D_\perp$ (circles) and $D_\parallel$ (diamonds), extrapolated to $1/H = 0$, are plotted as functions of the volume fraction. Note that, as in the experimental results, the values of two coefficients are close to each other.

Since no previous numerical results exist to which our data can be compared, the only possible comparison is with experimental findings but, no method currently exist for measuring the gradient diffusion coefficient directly. By fitting particle migration data to the solution of model equations, however, Leighton & Acrivos (1987*b*) were able to extract values for the gradient diffusivity at constant stress which they then fitted with the expression

$$D_1 = \frac{1}{3}\phi^2 \left(1 + \frac{1}{2}e^{8.8\phi}\right). \tag{5.25}$$

An alternative expression for the gradient diffusivity was given by Phillips *et al.* (1992),

$$D_2 = 0.65\,\phi^2 \frac{1}{\eta}\frac{\mathrm{d}\eta}{\mathrm{d}\phi}, \tag{5.26}$$

where $\eta$ is the effective relative viscosity of the suspension taken to be

$$\eta = \left(1 - \frac{\phi}{0.68}\right)^{-1.82}. \tag{5.27}$$

In figure 16, $D_1$ given by (5.25), is plotted as a solid line and $D_2$, as defined by equation (5.26), as a dashed line. Clearly both are in good agreement with the values for the diffusion coefficient as obtained from our numerical simulations and extrapolated to $1/H=0$.

## 6. Summary and concluding remarks

The flow of a monodisperse, neutrally buoyant, homogeneous suspension of non-Brownian solid spheres in simple shear, was simulated via Stokesian dynamics starting from a large number of independent hard-sphere configurations and ensemble averaging the results. We established that the particle displacements in the directions normal to the flow have Gaussian distributions the variances of which grow quadratically for small times, hence implying the absence of short-time self diffusion in contrast to Brownian suspensions, and linearly with time but only for time scales exceeding the correlation time which itself is quite large (shown in table 3). By extending the computations to longer time intervals than had been done previously, we were thereby able to compute the long-time self-diffusivities (shown in figure 9) for a suspension of non-Brownian particles subjected to shear flow. Our procedure, which also permitted us to evaluate the autocorrelation for the velocities in the directions normal to the flow, has the great advantage of breaking up the computational effort of a single long simulation into a number of more affordable short-lived independent simulations and is especially advantageous in computations using a parallel machine or a cluster of computers. We further confirmed that the integral of these velocity autocorrelation functions produce an alternate way of determining the self-diffusivities. We also found that, upon reversal of the direction of shear, the particles trace back their trajectory but only for a time that is shorter than the correlation time, after which they continue their diffusive motion as if the flow reversal had never occurred.

Finally, a new method, motivated by the work of Marchioro *et al.* (2000*a*,*b*), was developed for computing the gradient diffusion coefficients directly from numerical simulations. This method is based on artificially alterating the probability of an ensemble of configurations flowing under shear so that, according to the new probability density, defined in equation (5.3), the suspension becomes slightly inhomogeneous and a flux of particles is thereby generated. The balance of the different contributions to the particle-flux allows the calculation of the gradient diffusivity. To the best of our knowledge this is the first time that the shear-induced gradient diffusion has been determined directly either experimentally or via numerical computations.

Although all the results referred to above were found to apply even when $N$, the number of particles within the unit cell, was as low as 16, the numerical values of the diffusion coefficients become significantly larger when $N$ is increased (see tables 2 and 4), in contrast with the effective viscosity which can be computed accurately and independently of $N$ even if the latter is as small as 16 (see, e.g. Marchioro *et al.* 2000*a*). Consequently, for each volume fraction $\phi$ we evaluated these diffusion coefficients for different number of particles in the unit cell and then extrapolated the results to a cell of infinite extension and therefore containing an infinite number of particles (the results of this extrapolation are shown in table 5 and in figures 9 and 16). But since the errors associated with such extrapolations are, of course, unknown, it would be highly desirable if simulations could be performed with $N$ much larger than 64, the maximum number of particles per unit cell used in this study. This could only be accomplished, however, by employing more powerful computers or more efficient numerical algorithms and techniques than those currently available.

**Acknowledgments**

The authors wish to thank Dr. J. R. Melrose and Professor J. F. Brady for the use of their simulation codes and for important discussions regarding the oscillations of the intensity of the velocity fluctuations, professor B. Khusid for his constructive criticisms, and Dr. Y. Wang for his help during the early stage of this study.

This work was supported in part by the National Science Foundation under Grant CTS-9711442. This research was also supported in part by the NSF cooperative agreement ACI-9619020 through computing resources provided by the National Partnership for Advanced Computational Infrastructure at the Caltech Center for Advance Computing Research.

| shape of the box: $L \times H$ | frequency of the highest-peak beyond $\nu=0$ | magnitude of the peak beyond $\nu=0$ |
|---|---|---|
| $20 \times 10$ | 1/2 | $2\ 10^{-1}$ |
| $10 \times 10$ | 1 | $4\ 10^{-2}$ |
| $10 \times 15$ | 3/2 | $7\ 10^{-3}$ |
| $10 \times 20$ | 2 | $2\ 10^{-3}$ |

TABLE 1. Values and locations of the peaks of $F_{\langle vv \rangle}(\nu)$, the Fourier transform of $\langle v^2(\tau) \rangle$ defined in (4.4), in a monolayer, where $L$ and $H$ are, respectively, the extension of the cell in the $x$ and $y$ directions. The length of the box in the $z$ direction, $L_z$ was always set equal to 10 particle radii and the areal fraction was 50%.

| $N$ | $N_c$ | $1/H$ | $\hat{D}_{yy}$ | $\tilde{D}_{yy}$ | $\hat{D}_{zz}$ | $\tilde{D}_{zz}$ |
|---|---|---|---|---|---|---|
| 16 | 512 | 0.1735 | 0.019±0.003 | 0.019±0.003 | 0.009±0.001 | 0.008±0.001 |
| 27 | 256 | 0.1457 | 0.027±0.003 | 0.025±0.006 | 0.012±0.001 | 0.012±0.001 |
| 54 | 32 | 0.1157 | 0.037±0.012 | 0.038±0.020 | 0.014±0.006 | 0.012±0.008 |
| 64 | 32 | 0.1093 | 0.038±0.009 | 0.037±0.025 | 0.015±0.004 | 0.013±0.008 |
| $\infty$ | | 0 | 0.069 | 0.071 | 0.024 | 0.019 |

TABLE 2. Values of the self diffusivities evaluated for a volume fraction $\phi = 35\%$. $N$ is the number of particles in the unit cubic cell, $N_c$ is the number of configurations in the ensemble, and $H$ is the height of the cell. The values for $1/H=0$ were obtained by linear extrapolation.

| $N$ | $N_c$ | 15% | 25% | 35% | 45% |
|---|---|---|---|---|---|
| 16 | 512 | 8.6 | 5.4 | 4.1 | 2.9 |
| 27 | 256 | 6.5 | 5.3 | 3.5 | 3.0 |
| 54 | 32  | 8.5 | 5.0 | 3.2 | 3.1 |
| 64 | 32  | 6.5 | 4.8 | 4.7 | 3.2 |

TABLE 3. Values of the correlation time $T_c$, defined by equation (4.15), for $\phi$ =15%, 25%, 35%, and 45%, $N$ is the number of particles in the unit cell, $N_c$ the number of configurations in the ensemble.

| $N$ | $N_c$ | $1/H$ | $D_\parallel$ | $D_\perp$ |
|---|---|---|---|---|
| 16 | 512 | 0.1735 | 0.10±0.05 | 0.08 ± 0.03 |
| 27 | 256 | 0.1457 | 0.13±0.05 | 0.11 ± 0.02 |
| 54 | 128 | 0.1157 | 0.15±0.08 | 0.17 ± 0.02 |
| 64 | 128 | 0.1093 | 0.18±0.04 | 0.18 ± 0.02 |
| $\infty$ |  | 0 | 0.29 | 0.35 |

TABLE 4. Values of the gradient diffusivities evaluated for a volume fraction of $\phi = 35\%$, with $N$ number of particles in the unit cell, $N_c$ number of configurations in the ensemble, and $1/H$ is inverse of the cell size. The values for $1/H=0$ were obtained by linear extrapolation.

| $\phi$ | $\hat{D}_{yy}$ | $\tilde{D}_{yy}$ | $D_\parallel$ | $D_\perp$ |
|---|---|---|---|---|
| 15% | 0.0082 | 0.0033 | 0.10 | 0.083 |
| 25% | 0.027 | 0.010 | 0.14 | 0.13 |
| 35% | 0.069 | 0.071 | 0.29 | 0.35 |
| 35% | 0.078 | 0.050 | 0.54 | 1.16 |

TABLE 5. Summary of the values of the self and gradient diffusivities evaluated for different volume fractions, obtained by linear extrapolation to $1/H=0$.

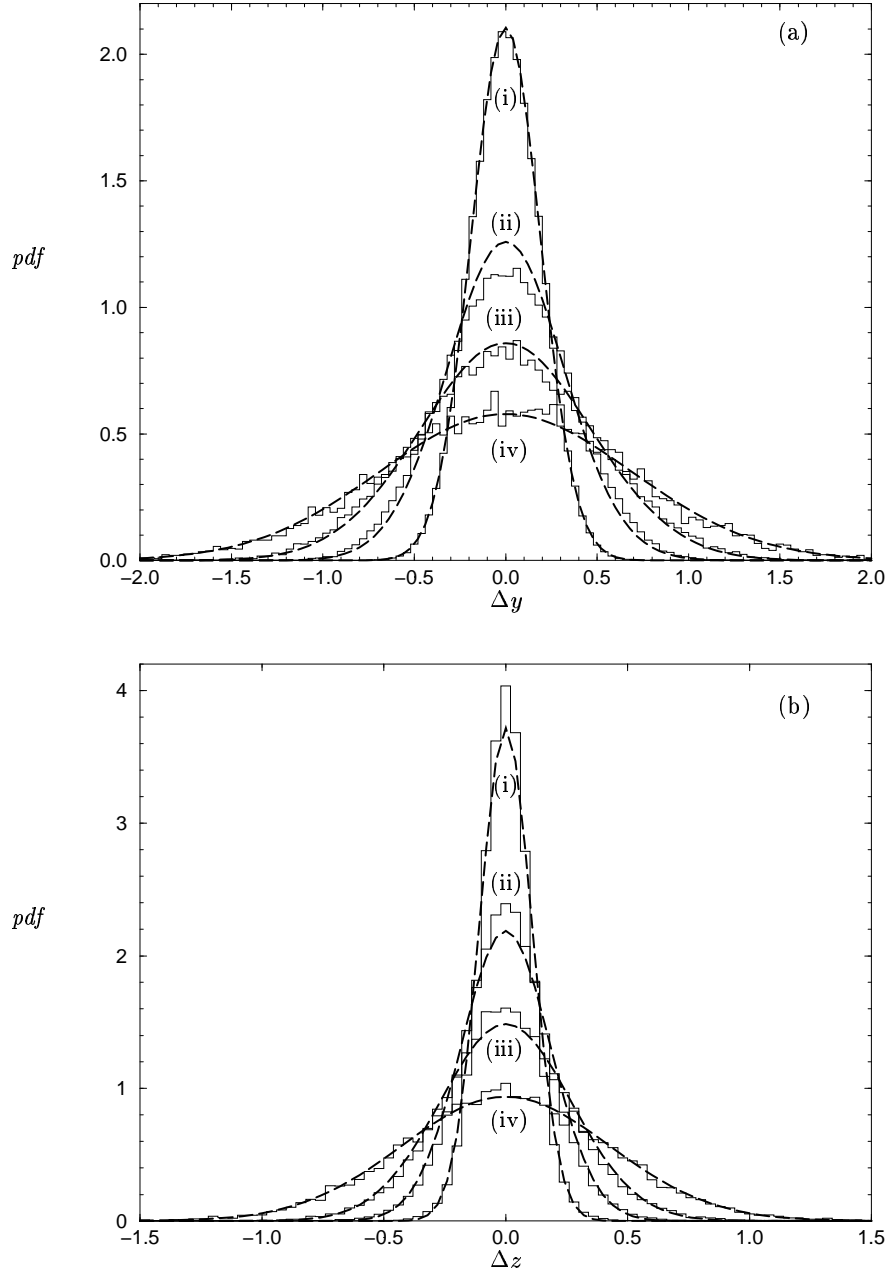

FIGURE 1. The probability density function (*pdf*) of (a) $\Delta y$ and (b) $\Delta z$, thin solid line, for $\tau=5$, $N = 16$, $N_c = 512$, and $\phi = 25\%$; $t=0.5$ (i), $t=1$ (ii), $t=2$ (iii), and $t=8$ (iv). The dashed lines are the density functions for a Gaussian distribution with zero mean and the same variance as the numerical data.

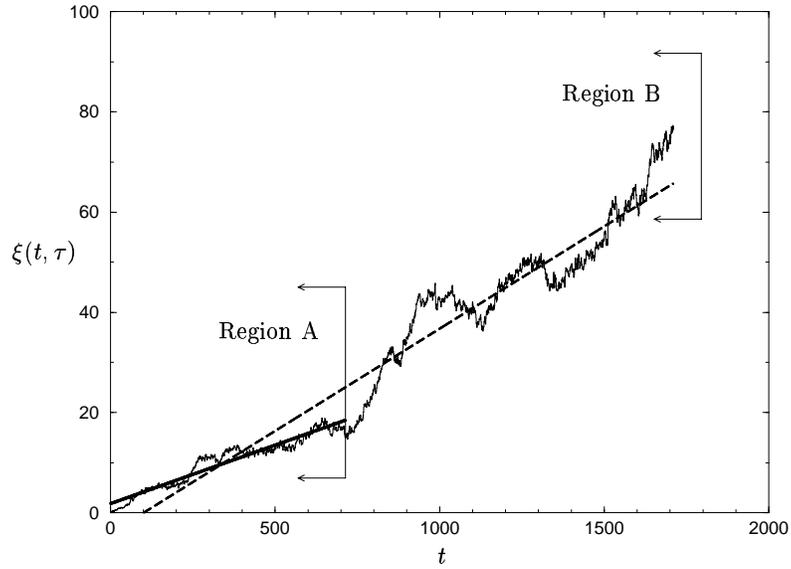

FIGURE 2. Plot of $\xi(t,\tau)$, as defined in (3.2), for $N = 27$, $\phi=35\%$. The two lines are linear least-squares fits in region A (thick solid line) and region B (dashed line).

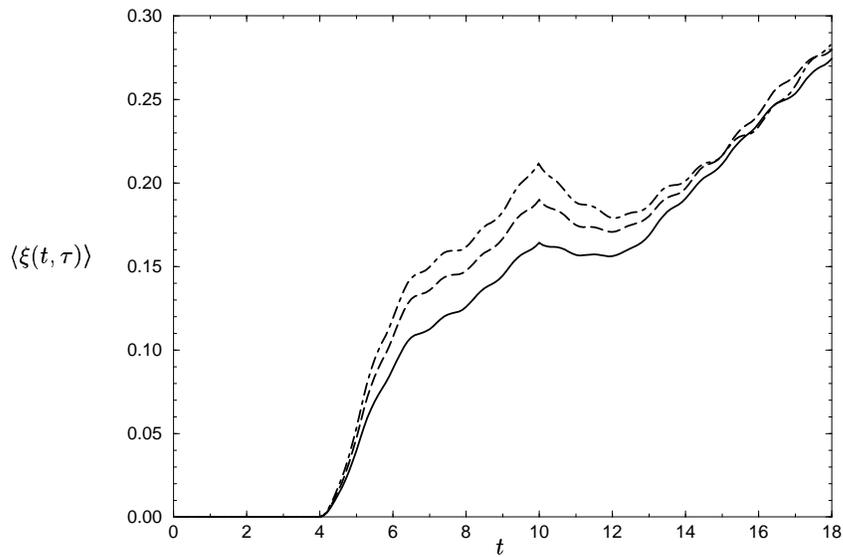

FIGURE 3. Plot of $\langle\xi(t,\tau)\rangle$ for $\tau=4$, $F_0=10$, $N_c=512$, $\lambda=10^3$ (solid line), $10^4$ (dashed line), $10^5$ (dot-dashed line). At time $t=T=6$ the direction of the shear is reversed.

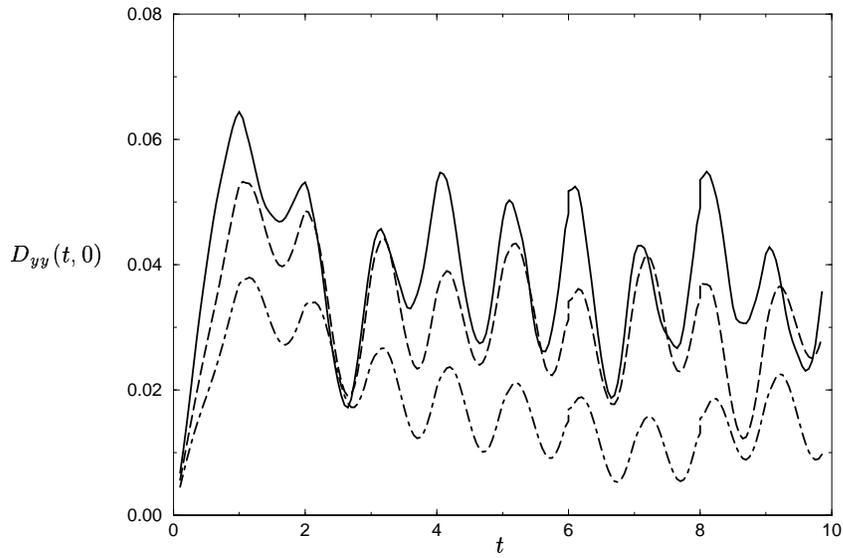

FIGURE 4. Graph of $D_{yy}(t,0)$ defined in equation (4.1) for $\phi = 25\%$ (dot-dashed line), $\phi = 35\%$ (dashed line), $\phi = 45\%$ (solid line), for $N=27$ and $N_c=512$.

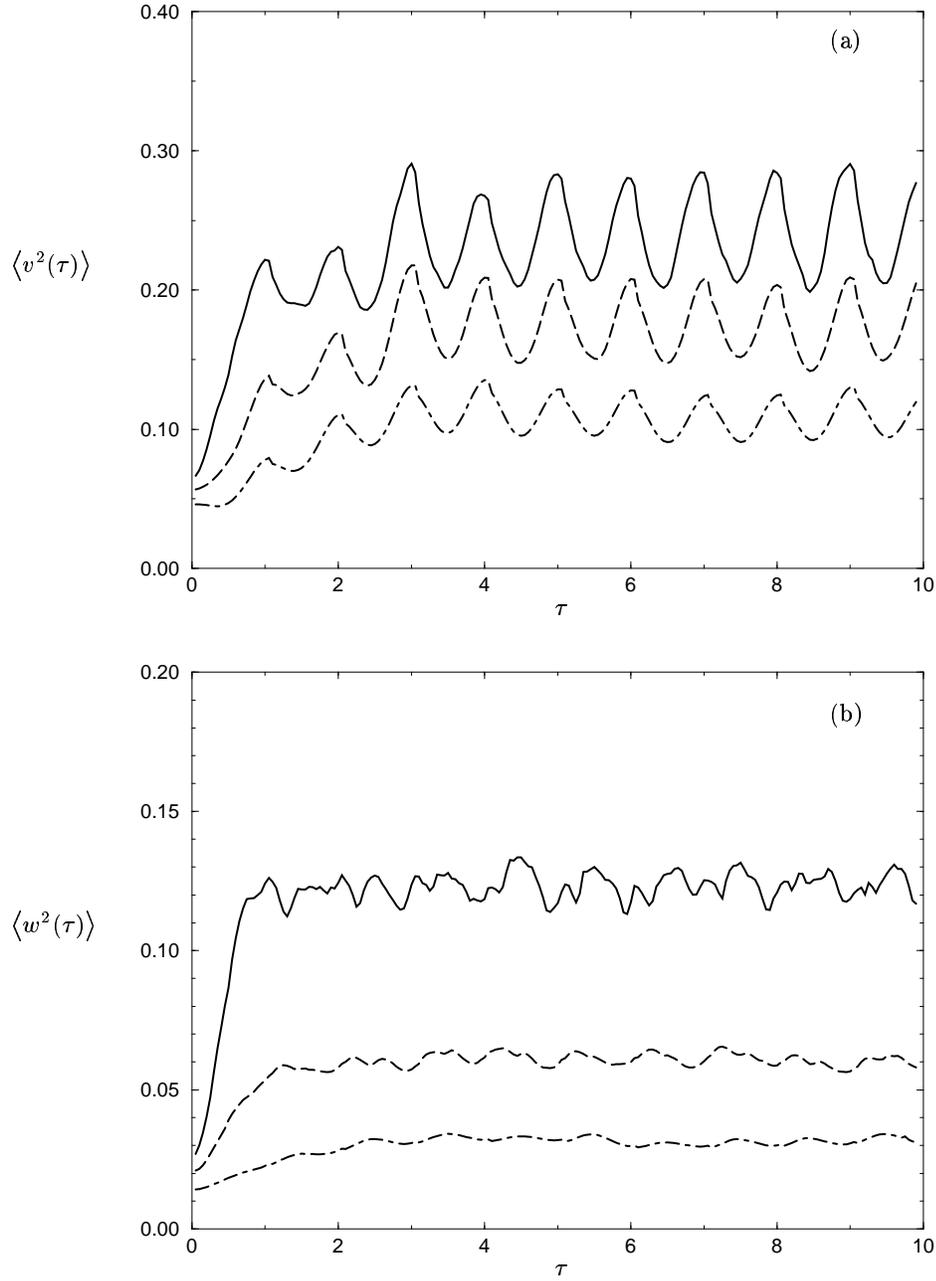

FIGURE 5. Graph of (a) $\langle v^2(\tau) \rangle$ defined in equation (4.3), and (b) $\langle w^2(\tau) \rangle$ for $\phi = 25\%$ (dot-dashed line), $\phi = 35\%$ (dashed line), $\phi = 45\%$ (solid line), for $N=27$ and $N_c=512$.

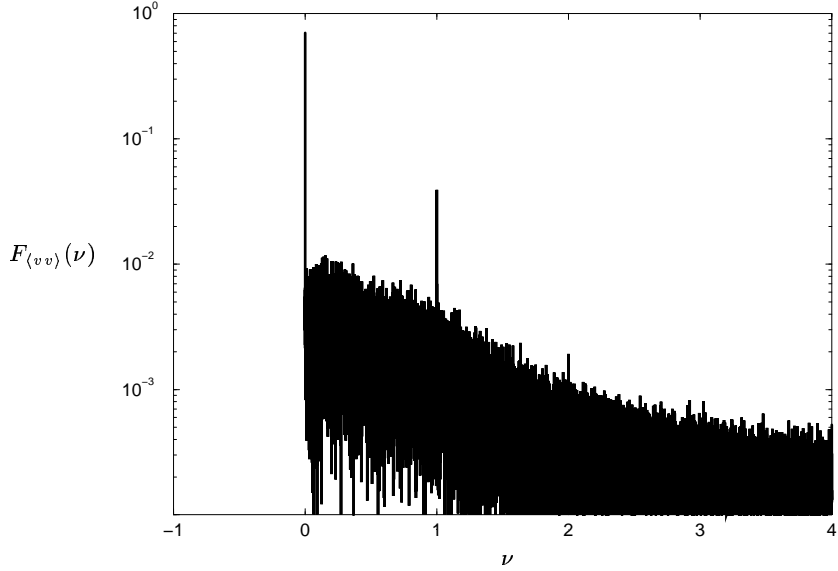

FIGURE 6. Plot of $F_{\langle vv \rangle}(\nu)$, the Fourier transform of $\langle v^2(\tau) \rangle$ defined in (4.4), for a cubic box with $N=16$, $N_c=1$, $\phi = 45\%$, and $\tau_\infty = 6554$.

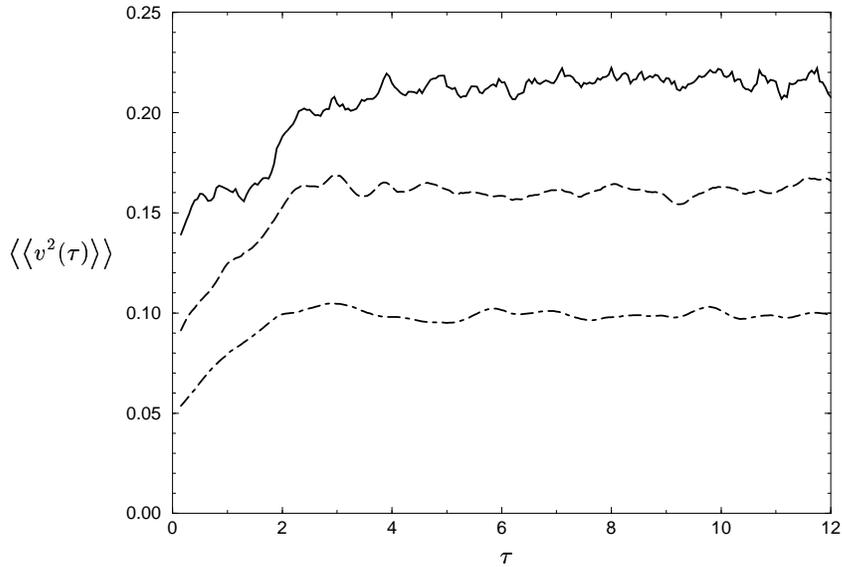

FIGURE 7. Graph of $\langle\langle v^2(\tau) \rangle\rangle$ as defined in equation (4.9), for $\phi = 25\%$ (dot-dashed line), $\phi = 35\%$ (dashed line), $\phi = 45\%$ (solid line), for $N=16$ and $N_c=512$.

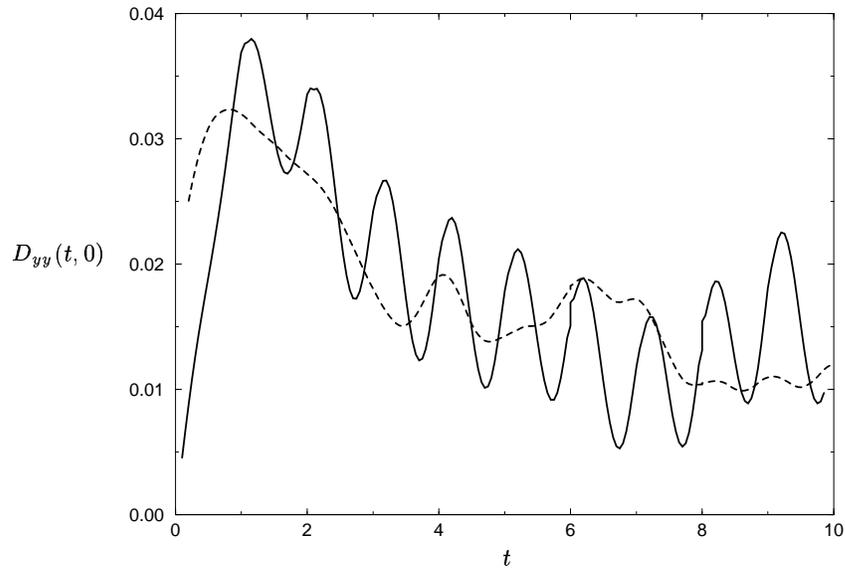

FIGURE 8. Graph of $D_{yy}(t,0)$, as defined in (4.1), where the average displacements are defined as in (4.2) (solid line) and in (4.10) (dashed line), $N = 27$, $N_c = 512$, $\phi = 15\%$.

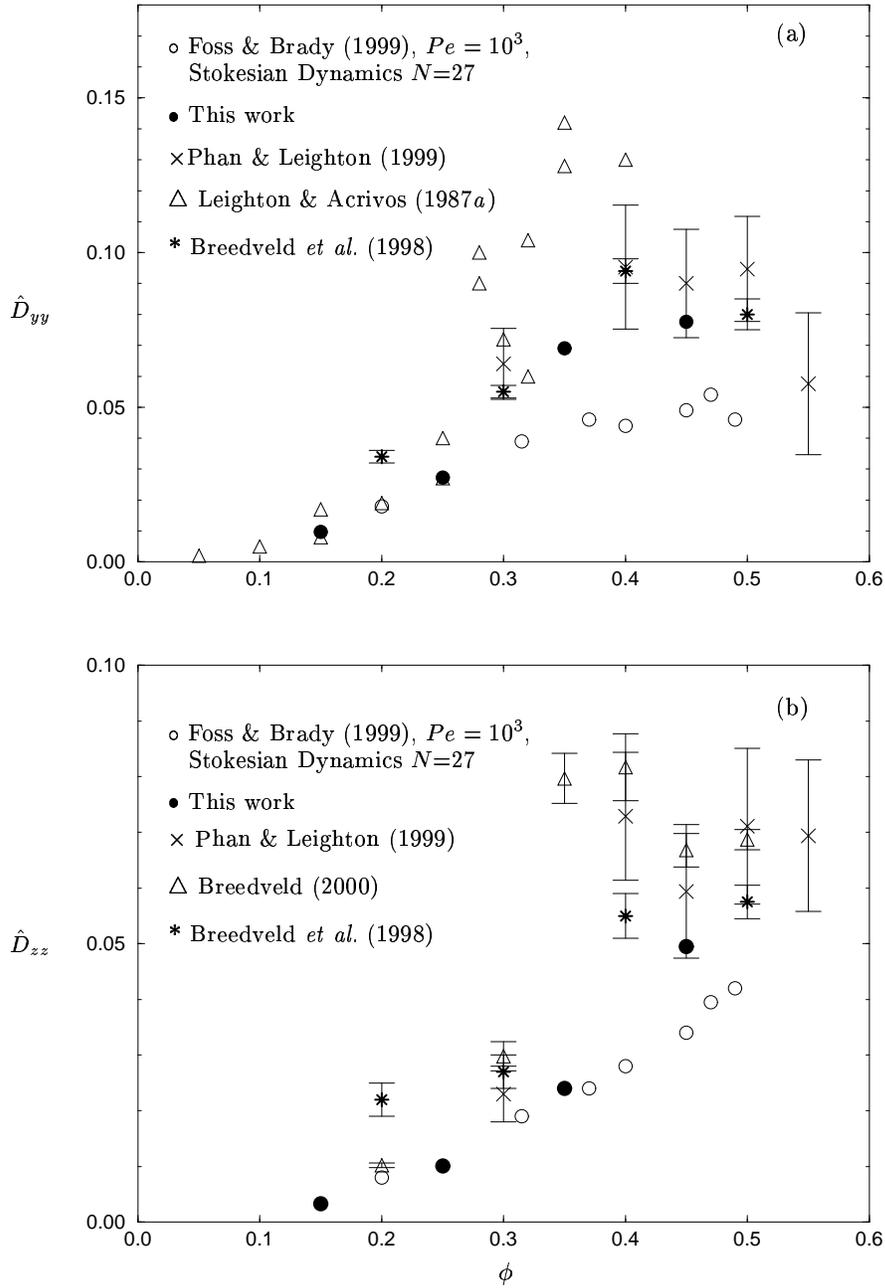

FIGURE 9. Comparison of the (a) $yy$-component and (b) $zz$-component of the long-time self-diffusion tensor extrapolated to $1/H=0$ with earlier experiment and computational results.

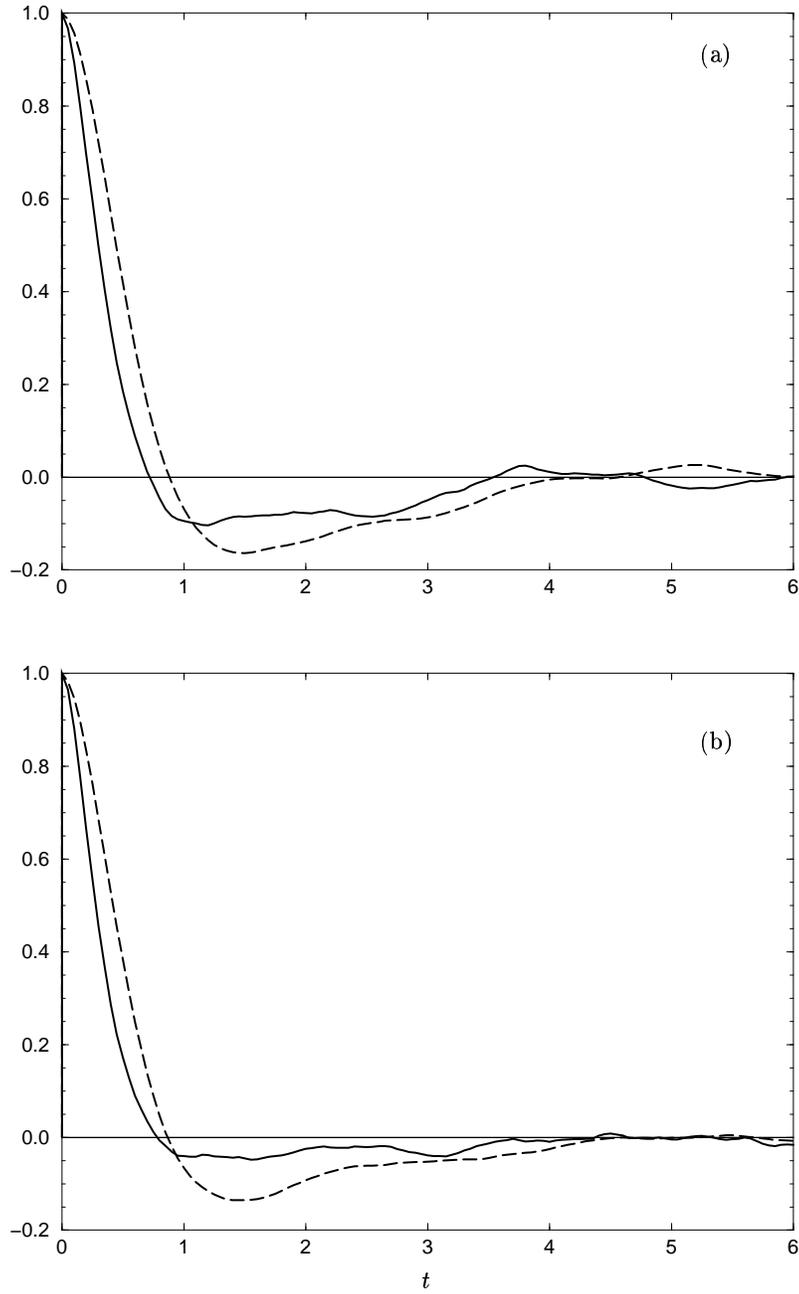

FIGURE 10. Autocorrelation functions for the velocity (a) $y$-component and (b) $z$-component, normalized by their respective values at $t=0$, for $\tau = 5$, $N=27$, $N_c=256$, $\phi = 35\%$ (dashed lines), and $\phi = 45\%$ (solid lines).

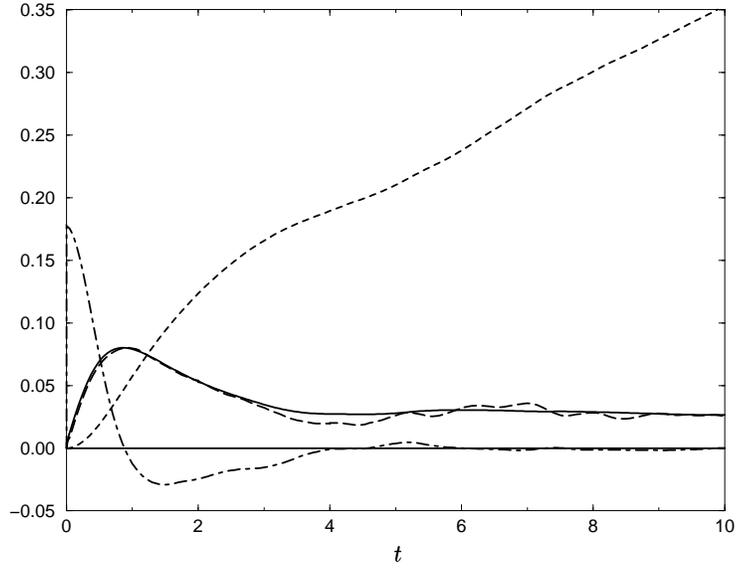

FIGURE 11. A plot of $\langle\langle v(t+\tau)v(\tau)\rangle\rangle$ (dot-dashed line), together with its integral (solid line), also $D_{yy}(t,\tau)$ (dashed line), and $\frac{1}{2}\langle\langle [y(t+\tau)-y(\tau)]^2\rangle\rangle$ (dotted line), for $N_c{=}512$, $N{=}27$, $\phi{=}35\%$, $\tau{=}5$.

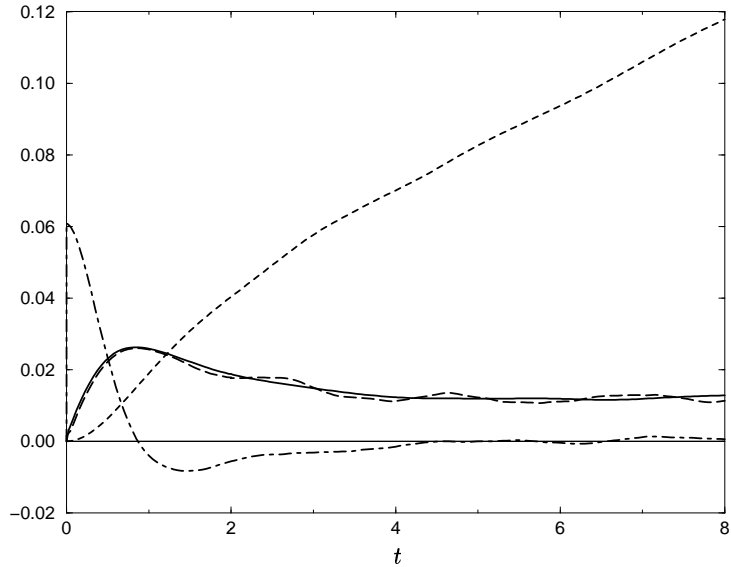

FIGURE 12. A plot of $\langle\langle w(t+\tau)w(\tau)\rangle\rangle$ (dot-dashed line), together with its integral (solid line), also $D_{zz}(t,\tau)$ (dashed line), and $\frac{1}{2}\langle\langle [z(t+\tau)-z(\tau)]^2\rangle\rangle$ (dotted line), for $N_c{=}512$, $N{=}27$, $\phi{=}35\%$, $\tau{=}5$.

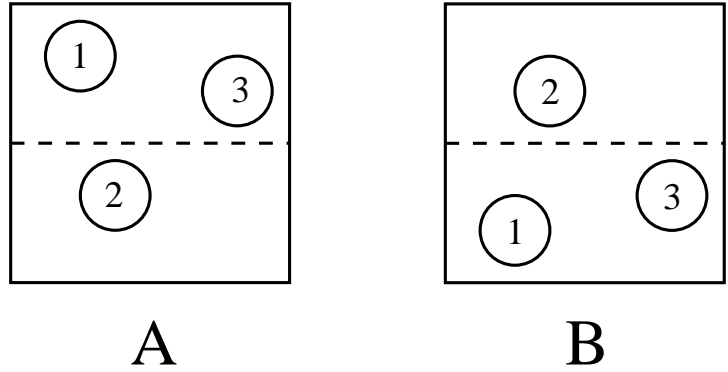

FIGURE 13. Two equally likely configurations in the homogeneous ensemble. Configuration B is obtained from A by a mirror symmetry about the centerline.

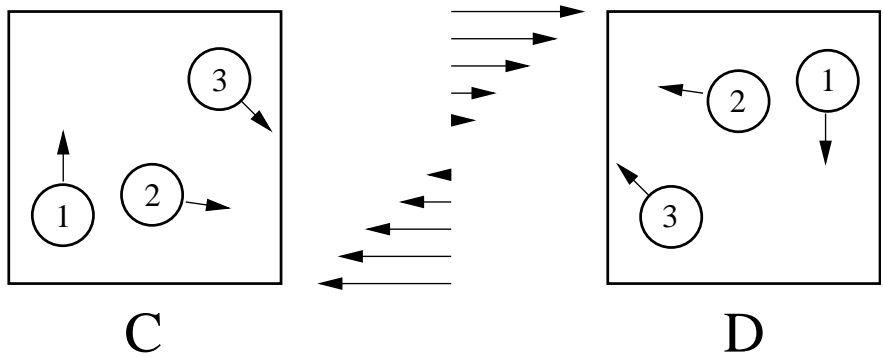

FIGURE 14. Two configurations that have the same probability in the homogeneous ensemble. Configuration D can be obtained from C by a rotation of 180° degrees. Notice how the velocity of the particles in the vertical direction are opposite to each other in the two configurations.

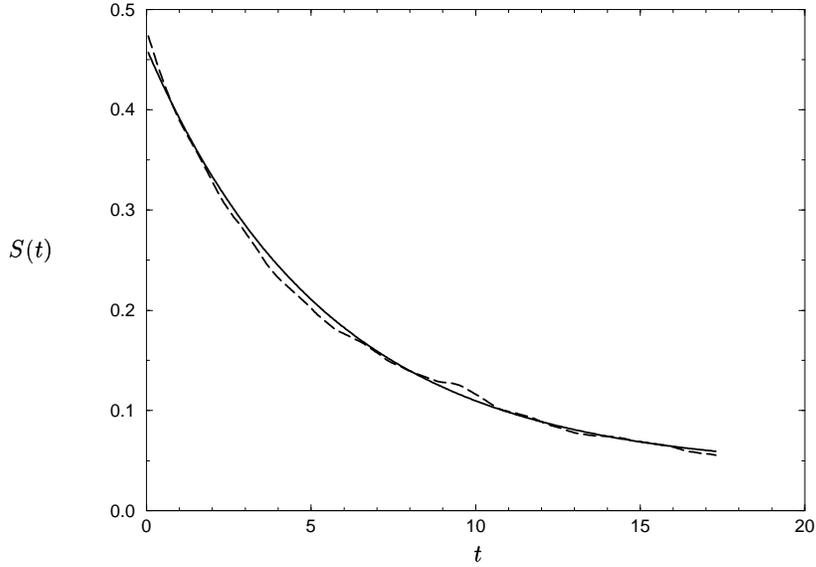

FIGURE 15. Computed values of $S(t)$ (dashed line) for $N_c=256$, $N=27$, $k=0.9956$, and a volume fraction $\phi=35\%$. The solid line, $y = 0.041 + 0.42\,e^{-0.181\,t}$, is a curve fit of $S(t)$ according to (5.24).

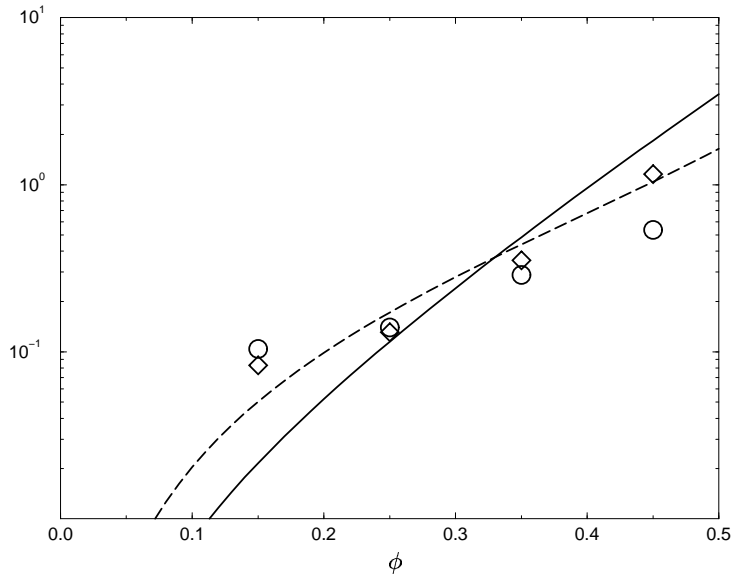

FIGURE 16. Values of $D_\parallel$ (circles) and $D_\perp$ (diamonds) extrapolated to $1/H = 0$ as functions of the volume fraction $\phi$. The solid line is $D_1$ as defined in equation (5.25) and the dashed line is $D_2$ as defined in equation (5.26).

**Table 1.** Values and locations of the peaks of $F_{\langle vv \rangle}(\nu)$, the Fourier transform of $\langle v^2(\tau) \rangle$ defined in (4.4), in a monolayer, where $L$ and $H$ are, respectively, the extension of the cell in the $x$ and $y$ directions. The length of the box in the $z$ direction, $L_z$ was always set equal to 10 particle radii and the areal fraction was 50%.

**Table 2.** Values of the self diffusivities evaluated for a volume fraction $\phi = 35\%$. $N$ is the number of particles in the unit cubic cell, $N_c$ is the number of configurations in the ensemble, and $H$ is the height of the cell. The values for $1/H=0$ were obtained by linear extrapolation.

**Table 3.** Values of the correlation time $T_c$, defined by equation (4.15), for $\phi =15\%$, 25%, 35%, and 45%, $N$ is the number of particles in the unit cell, $N_c$ the number of configurations in the ensemble.

**Table 4.** Values of the gradient diffusivities evaluated for a volume fraction of $\phi = 35\%$, with $N$ number of particles in the unit cell, $N_c$ number of configurations in the ensemble, and $1/H$ is inverse of the cell size. The values for $1/H=0$ were obtained by linear extrapolation.

**Table 5.** Summary of the values of the self and gradient diffusivities evaluated for different volume fractions, obtained by linear extrapolation to $1/H=0$.

**Figure captions**

**Figure 1.** The probability density function (*pdf*) of (a) $\Delta y$ and (b) $\Delta z$, thin solid line, for $\tau=5$, $N=16$, $N_c=512$, and $\phi=25\%$; $t=0.5$ (i), $t=1$ (ii), $t=2$ (iii), and $t=8$ (iv). The dashed lines are the density functions for a Gaussian distribution with zero mean and the same variance as the numerical data.

**Figure 2.** Plot of $\xi(t,\tau)$, as defined in (3.2), for $N=27$, $\phi=35\%$. The two lines are linear least-squares fits in region A (thick solid line) and region B (dashed line).

**Figure 3.** Plot of $\langle \xi(t,\tau)\rangle$ for $\tau=4$, $F_0=10$, $N_c=512$, $\lambda=10^3$ (solid line), $10^4$ (dashed line), $10^5$ (dot-dashed line). At time $t=T=6$ the direction of the shear is reversed.

**Figure 4.** Graph of $D_{yy}(t,0)$ defined in equation (4.1) for $\phi=25\%$ (dot-dashed line), $\phi=35\%$ (dashed line), $\phi=45\%$ (solid line), for $N=27$ and $N_c=512$.

**Figure 5.** Graph of (a) $\langle v^2(\tau)\rangle$ defined in equation (4.3), and (b) $\langle w^2(\tau)\rangle$ for $\phi=25\%$ (dot-dashed line), $\phi=35\%$ (dashed line), $\phi=45\%$ (solid line), for $N=27$ and $N_c=512$.

**Figure 6.** Plot of $F_{\langle vv\rangle}(\nu)$, the Fourier transform of $\langle v^2(\tau)\rangle$ defined in (4.4), for a cubic box with $N=16$, $N_c=1$, $\phi=45\%$, and $\tau_\infty=6554$.

**Figure 7.** Graph of $\langle\langle v^2(\tau)\rangle\rangle$ as defined in equation (4.9), for $\phi=25\%$ (dot-dashed line), $\phi=35\%$ (dashed line), $\phi=45\%$ (solid line), for $N=16$ and $N_c=512$.

**Figure 8.** Graph of $D_{yy}(t,0)$, as defined in (4.1), where the average displacements are defined as in (4.2) (solid line) and in (4.10) (dashed line), $N=27$, $N_c=512$, $\phi=15\%$.

**Figure 9.** Comparison of the (a) $yy$-component and (b) $zz$-component of the long-time self-diffusion tensor extrapolated to $1/H=0$ with earlier experiment and computational results.

**Figure 10.** Autocorrelation functions for the velocity (a) $y$-component and (b) $z$-

component, normalized by their respective values at $t=0$, for $\tau = 5$, $N=27$, $\phi = 35\%$ (dashed lines), and $\phi = 45\%$ (solid lines).

**Figure 11.** A plot of $\langle\langle v(t+\tau)v(\tau)\rangle\rangle$ (dot-dashed line), together with its integral (solid line), also $D_{yy}(t,\tau)$ (dashed line), and $\frac{1}{2}\langle\langle [y(t+\tau) - y(\tau)]^2\rangle\rangle$ (dotted line), for $N_c=512$, $N=27$, $\phi=35\%$, $\tau=5$.

**Figure 12.** A plot of $\langle\langle w(t+\tau)w(\tau)\rangle\rangle$ (dot-dashed line), together with its integral (solid line), also $D_{zz}(t,\tau)$ (dashed line), and $\frac{1}{2}\langle\langle [z(t+\tau) - z(\tau)]^2\rangle\rangle$ (dotted line), for $N_c=512$, $N=27$, $\phi=35\%$, $\tau=5$.

**Figure 13.** Two equally likely configurations in the homogeneous ensemble. Configuration B is obtained from A by a mirror symmetry about the centerline.

**Figure 14.** Two configurations that have the same probability in the homogeneous ensemble. Configuration D can be obtained from C by a rotation of 180° degrees. Notice how the velocity of the particles in the vertical direction are opposite to each other in the two configurations.

**Figure 15.** Computed values of $S(t)$ (dashed line) for $N_c=256$, $N=27$, $k=0.9956$, and a volume fraction $\phi=35\%$. The solid line, $y = 0.041 + 0.42\,e^{-0.181 t}$, is a curve fit of $S(t)$ according to (5.24).

**Figure 16.** Values of $D_\parallel$ (circles) and $D_\perp$ (diamonds) extrapolated to $1/H = 0$ as functions of the volume fraction $\phi$. The solid line is $D_1$ as defined in equation (5.25) and the dashed line is $D_2$ as defined in equation (5.26).